\documentclass[aps,prl,twocolumn,showpacs,groupedaddress]{revtex4}

\usepackage{amsmath}
\usepackage{graphicx}
\usepackage{dcolumn}
\usepackage{bm}

\begin{document}

\preprint{QF-HE32D}

\title{Zero-temperature equation of state of two-dimensional $^{\bm 3}$He}

\author{V. Grau}
\altaffiliation[Present address: ]{Departament d'Enginyeria Mec\`anica,
Universitat Polit\`ecnica de Catalunya}
\author{J. Boronat}
\author{J. Casulleras}
%\email{jordi.boronat@upc.es}
\affiliation{Departament de F\'\i sica i Enginyeria Nuclear, Campus Nord
B4-B5, Universitat Polit\`ecnica de Catalunya, E-08034: Barcelona, Spain}

\date{\today}

\begin{abstract}   
The equation of state of two-dimensional $^3$He at zero temperature has
been calculated using the  diffusion Monte Carlo method. By means of a
combination of  the fixed-node and released-node techniques it is shown
that  backflow correlations provide a very accurate equation of state.  
The results prove unambiguously the non-self-bound character of
two-dimensional $^3$He due to its Fermi statistics. We present 
solid evidence that the gas phase, predicted for the two-dimensional
system, can be extrapolated to the case of $^3$He adsorbed on a strong
substrate like graphite.

\end{abstract} 


\maketitle

$^3$He adsorbed on strongly interacting substrates like graphite, or on
top of bulk $^4$He or $^4$He films, constitute experimental realizations of
quasi-two dimensional Fermi systems. In the last
decades there has been a continued experimental effort to unveil
the fascinating properties of such a nearly perfect two-dimensional
Fermi liquid. Among these unique features, of  particular
relevance is the possibility of continuously increasing  the
areal density from an almost ideal gas behavior up to a strongly
correlated regime. This is the experimental situation observed for example
in the two first layers of $^3$He adsorbed on graphite. 
These experimental
findings indicate the non-existence of a self-bound $^3$He system.
In contrast, 
Cs\'athy \textit{et al.}~\cite{chan} have recently studied submonolayer
$^3$He-$^4$He mixture films on H$_2$ and claim that $^3$He atoms appear
to have condensed into a 2D self-bound liquid. Also, a recent theoretical study
of mixture films points out the possibility of a dimerized $^3$He phase in a
strictly 2D geometry~\cite{kromix}.
In fact, the question of a self-bound 2D 
$^3$He phase has been discussed for a 
long time both from 
experimental~\cite{morhard2,casey,bauerle,ho}
and theoretical
perspectives~\cite{novaco,nosanow,brami,clements2d,chester2d}.

Theoretical calculations concerning the 2D
$^3$He system and the  $^3$He films  are scarce in comparison with the 
corresponding ones for the boson isotope $^4$He. In addition to dealing
with a strongly correlated system like helium, the Fermi
statistics of $^3$He must be taken into account.
In one of the
pioneering works of the field, Novaco and Campbell~\cite{novaco}
calculated the equation of state of $^3$He
adsorbed on graphite. Using lowest-order Fermi corrections, they
concluded that the $^3$He film is in a gas state, contrary to
$^4$He which exhibits a well-established self-bound
character~\cite{whitlock,boro2d}. A
comparative study of bosons and fermions in 2D was performed by Miller
and Nosanow~\cite{nosanow} using the variational method. According to
their  approach, and using a Wu-Feenberg
expansion~\cite{wu} 
at lowest order, $^3$He cannot condense in
2D. More recently, Brami \textit{et al.}~\cite{brami} calculated the
properties of a 2D $^3$He film using variational Monte Carlo (VMC). 
They concluded that the
presence of a transverse degree of freedom, not present in two dimensions, 
allows
the system to gain enough additional binding energy to guarantee a
liquid phase with a very small energy. 
However, a recent Green's function Monte Carlo (GFMC) calculation 
by Whitlock \textit{et al.}~\cite{whitlock2}, 
of a $^4$He film adsorbed on graphite, 
has shown that the energy gain with respect to
the ideal 2D system is much smaller than the one estimated in
Ref. \cite{brami}.

In this work, we use the diffusion Monte Carlo (DMC) method,  
which for  bulk 
$^3$He has produced, for the first time, close quantitative
agreement between theoretical and experimental results for the equation of
state~\cite{bulk3}.
Our aim is to achieve the same accuracy in the present study of a 
strictly 2D $^3$He fluid. The energies
obtained constitute upper bounds to the eigenvalues of the many-body
Schr\"odinger equation, but the method can measure  the \textit{quality} of
the bounds and provide a means of improving them. The results presented
in this work show that backflow correlations are 
sufficient to bring the
systematic error to the  level of the statistical noise.
The resulting equation of state implies the non-existence of
self-binding,  in agreement with most experimental observations.

%The zero-temperature equation of state of two-dimensional $^3$He has been
%calculated using the DMC method. 
DMC \cite{anderson} is a stochastic method that solves the $N$-body
imaginary-time Schr\"odinger equation
%\begin{eqnarray}
%-\frac{\partial f({\bm R},t)}{\partial t}  & =  &  -D\,
%{\bm \nabla}^2_{{\bm R}}
%f({\bm R},t)+D\,{\bm \nabla}_{{\bm R}} \left( {\bm F}({\bm R})
%\,f({\bm R},t)\,
%\right)  \nonumber \\ 
%& & + \left(E_{\text L}({\bm R})-E \right)\,f({\bm R},t) \ ,
%\label{dmc1}                                        
%\end{eqnarray}
for the wave function $f({\bm R},t)= \psi({\bm R})\,\Psi({\bm R},t)$, with 
$\psi({\bm R})$ a trial wave function used for importance sampling (see
Ref. \cite{boro4he} for a more detailed description of the actual DMC
algorithm used). 
%In Eq.
%\ref{dmc1}, $D=\hbar^2 /(2m)$, $E_{\text L}({\bm R})=\psi({\bm R})^{-1} H 
%\psi({\bm R})$ is the local energy, and ${\bm F}({\bm R}) = 2\, 
%\psi({\bm R})^{-1} {\bm \nabla}_{{\bm R}} \psi({\bm R})$ is the quantum or
%drift force. The probability density $f({\bm R},t)$, which is represented by
%a set of walkers ${\bm R}=\{ {\bm r}_1,\ldots,{\bm r}_N \}$, evolves
%according to a short-time Green function accurate to $O((\Delta
%t)^3)$~\cite{boro4he}. With
%this prescription, the energy of the system shows a systematic error that
%grows like $(\Delta t)^2$ and thus the time-step dependence is practically
%eliminated below a characteristic $\Delta t$ value.
The first and simplest approximation for the trial wave function $\psi$ is
the Jastrow-Slater form
\begin{equation}
\psi^{\text F} = \psi_{\text J} \, D^\uparrow D^\downarrow \ ,
\label{feynman}
\end{equation}
with a Jastrow factor $\psi_{\text J}=\prod_{i<j}^N \exp (u(r_{ij}))$ 
accounting for the dynamical
correlations induced by the interatomic potential, and $D^\uparrow$
($D^\downarrow$) a plane-wave Slater determinant for the spin-up (spin-down) 
atoms. 

The nodal surface provided by $\psi^{\text F}$ corresponds to the 2D
free Fermi gas. This is a first  approximation since 
the real nodal surface is  modified by   
dynamical correlations. This influence is  contained in the
time-dependent Schr\"odinger equation. Starting with $\psi^{\text F}$ as
zeroth order, a straightforward calculation shows that the first order 
correction to the wave
function incorporates the so called backflow correlations, a name 
used  in analogy to
the same type of corrections introduced by Feynman and Cohen~\cite{feyco} 
in their famous
paper on the $^4$He phonon-roton spectrum. In this new wave function
$\psi^{\text{BF}}$ the arguments of the orbitals $\varphi_\alpha(i)$ 
entering $D^\uparrow$ and 
$D^\downarrow$ are shifted under the influence of the medium,
\begin{equation}
\exp{ \{ i {\bm k}_\alpha \cdot \tilde{{\bm r}}_i \} } \equiv 
\exp{ \left\{ i {\bm k}_\alpha  \cdot \left(  {\bm r}_i + 
\lambda_{\text B} \sum_{j \neq i}^N
\eta(r_{ij}) {\bm r}_{ij} \right) \right\} }  
%\hspace{0.5cm} (i=1,\ldots,N)
\ .
\label{cohen}
\end{equation}  

The two-body correlation factor is of McMillan type, $u(r)=-0.5 (b/r)^5$, and
the backflow function is approximated by a gaussian, $\eta(r)=\exp [
-( (r-r_{\text B})/\omega_{\text B})^2 ]$. The parameters of $\psi$ have
been optimized using VMC; the optimal values are
$b=1.16\sigma$, $\lambda_{\text B}= 0.40$, $r_{\text B}= 0.75\sigma$, and
$\omega_{\text B} = 0.54\sigma$ ($\sigma=2.556$\AA). The density dependence
of this set of parameters in the region  studied here is very weak and can
be neglected. The interatomic interaction corresponds to the HFD-B(HE)
potential from Aziz \textit{et al.}~\cite{aziz}; its use in bulk
$^4$He~\cite{boro4he} and $^3$He~\cite{bulk3}
has allowed for a very accurate calculation of their respective equations
of state.

%In order to ensure the positiveness of the wave function
%$f({\bm R},t)$, in a fermion calculation like the present one, 
%the fixed-node (FN) approximation is used. 

\begin{figure}
\centering
\includegraphics[width=7cm]{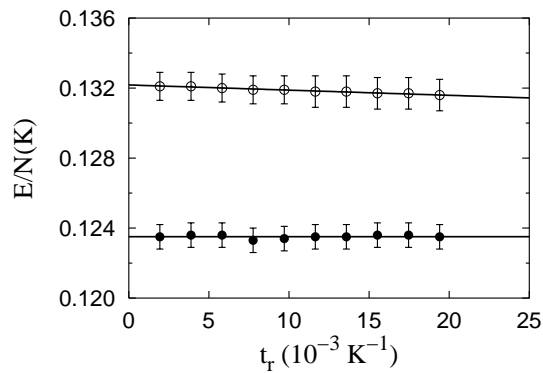}
\caption{\label{fig:fig1} RN energies as a function of the released time
$t_{\text r}$. Empty and filled circles stand for $\psi^{\text F}$ and 
$\psi^{\text{BF}}$ as trial wave functions, respectively. Solid lines are
linear fits to the calculated data.}
\end{figure}

The upper-bound to the energy  
provided by the fixed-node (FN) approximation depends on the accuracy 
of the nodal surface of $\psi({\bm R})$;  if it is the exact one then 
FN generates the eigenvalue. 
%of Eq.~(\ref{dmc1}).
Otherwise, FN yields a variational estimate of the energy of the system,
but provides no indication of the accuracy of this estimate. 
In a preceding work on bulk $^3$He~\cite{bulk3} we devised a
combined method that incorporates FN, the released-node (RN) method,
and an analytical prescription to improve $\psi({\bm R})$. The RN method
does not generally yield the exact energy, due to the growth of the boson
component,  but the initial slope of the energy vs. the released time is
readily accessible. From this slope one can guess the difference between
the FN energy and the eigenvalue; an exact wave function generates zero
slope. In Fig. 1, RN results at medium density ($\rho=0.10\sigma^{-2}$) are
shown for the two trial wave functions $\psi^{\text F}$ and
$\psi^{\text{BF}}$ 
reported above. The decrease of the energy going from $\psi^{\text F}$ to  
$\psi^{\text{BF}}$ is $\sim 0.01$ K, a small figure in absolute value
but of the same relative magnitude as in bulk. More importantly, 
the RN energies shown
in Fig. 1 illustrate the different behavior of $E/N$ with the
released time $t_{\text r}$: the $\psi^{\text F}$ results show a linear
decreasing trend which disappears when backflow correlations are
introduced. 
In fact, a fit to the slope of $E/N(\psi^{\text{BF}})$ gives a value
compatible with zero. This remarkable result can be combined with an
estimation of the released time from $E/N(\psi^{\text{F}})$ which indicates
that the systematic error of the BF energies due to the nodal surface
is in the mK order of magnitude, \textit{i.e.}, 
it has been brought to the level of
the typical statistical errors of this work. 
Additional insight on the high quality of the nodal
surface provided by the inclusion of backflow correlations is  reached by
incorporating in $\psi$ the next-order analytical terms~\cite{bulk3}. 
The results obtained with the new $\psi$, which includes explicit
three-body correlations in the backflow operator, 
do not show any improvements with respect to the BF 
energies. Therefore, the
DMC results obtained for 2D $^3$He, using the optimized backflow
correlations, are essentially exact; it is worth noticing that the same
method gives  unprecedent accuracy in the calculation of
the bulk $^3$He equation of state.

\begin{table}
\caption{\label{tab:table1}DMC total and partial energies (in K) of 2D
$^3$He. The last column shows the upper bound  $(E/N)^{\text F}$, relative
to the DMC energies (col. 1).  
Figures in parenthesis are the statistical errors. }
%is the difference between the   
%$\psi^{\text{BF}}$ and $\psi^{\text F}$ energies, $\Delta E/N=
%(E/N)^{\text{
%BF}} - (E/N)^{\text F}$. Figures in parenthesis are the statistical errors. }
\begin{ruledtabular}
\begin{tabular}{ccccc}
%Left\footnote{Note a.}&Centered\footnote{Note b.}&Right\\
%\hline
%          & \multicolumn{2}{c}{$\psi^{\text F}$} &
%	  \multicolumn{2}{c}{$\psi^{\text BF}$}  \\ 
%\hline
$\rho(\sigma^{-2})$  &  $E/N$    &  $T/N$  &  $V/N$   &  $\Delta E/N$     \\ 
\hline
0.01     &  0.0262(4)   &  0.0884(11) &  -0.0622(11)  &   0.003   \\ 
0.06     &  0.0971(26)  &  0.6678(30) &  -0.5707(30)  &   0.003  \\
0.10     &  0.1244(18)  &  1.2015(70) &  -1.0771(70)  &   0.008  \\
0.17     &  0.2204(22)  &  2.4329(87) &  -2.2125(87)  &   0.012  \\
0.23     &  0.3939(22)  &  3.7414(87) &  -3.3475(87)  &   0.038  \\
\end{tabular}
\end{ruledtabular}
\end{table}    

Results for the total and partial energies of the 2D $^3$He system as a
function of the surface density are reported in Table I. The potential
energies per particle have been obtained using a pure estimation method
~\cite{pures} 
in order to avoid any
biases coming from the trial wave function. The kinetic energy comes from
the difference between the total and potential energies. All the
calculations have been carried out with 90 atoms, the finite-size
simulation effects having been corrected for, and in practice eliminated, by summing
up energy-tail contributions coming  from both the dynamical and
statistical parts. The final DMC energies are
positive for any density and result from a significant cancellation between $T/N$
and $V/N$, a usual feature in condensed helium. The last column in Table I
contains the decrease of the energy when the nodal
surface is improved by the inclusion of backflow correlations. As could
be expected, $\Delta E/N$ increases  with $\rho$ as does the relevance of
correlations. 

\begin{figure}
\centering
\includegraphics[width=6.8cm]{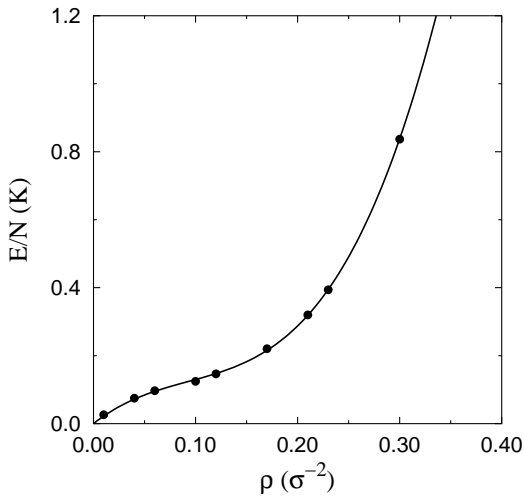}
\caption{\label{fig:fig2} Energy per particle of 2D $^3$He as a function of
the density. The statistical error bars are smaller than the size of the
symbols. The solid line corresponds to a third-degree polynomial fit
(\protect\ref{eqestatpol}) to the DMC data. 
}
\end{figure}
 
The equation of state of 2D $^3$He is shown in Fig. 2 for the range of
densities  studied. In this region, our results are well parameterized
by a cubic polynomial (solid line)
\begin{equation}
(E/N) (\rho) =  a \rho + b  \rho^2 + c \rho^3   \ ,
\label{eqestatpol} 
\end{equation}
with optimal parameters $a=2.376(74)$K$\sigma^2$, $b=
-16.87(81)$K$\sigma^4$, and $c=0.608(21)\cdot 10^2$K$\sigma^6$. At very
small densities ($\rho \alt 0.05\sigma^{-2}$), the energy grows linearly
as in a free Fermi gas ($E_{\text F}/N =\hbar^2/(2m) \, \pi \rho$)
but with different slope. At medium densities, there is a
clear change in the slope with a flatter region suggesting the emergence of
a minimum. Nevertheless, this minimum does not appear and the energy
remains always positive. For $\rho \agt 0.25\sigma^{-2}$ the energy
increases much faster as the density approaches the freezing point,
 which experimentally
is observed at $\rho \simeq 0.394\sigma^{-2}$~\cite{bauerle}.

\begin{figure}
\centering
\includegraphics[width=6.8cm]{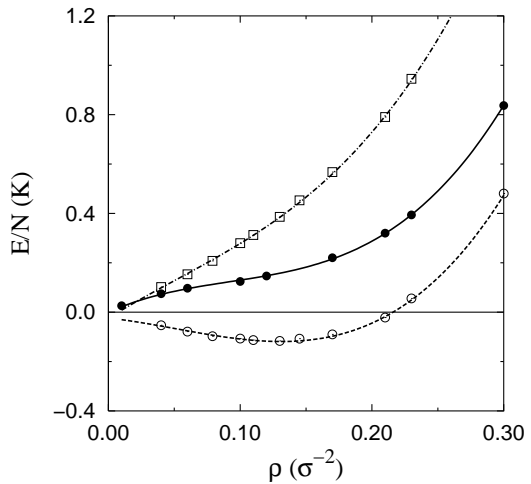}
\caption{\label{fig:fig3} Influence of the Fermi statistics on the energy
of 2D $^3$He. Filled and empty circles correspond to Fermi and 
 Bose $^3$He, respectively. Squares  represent the sum of the Boson
 energy and the Fermi gas kinetic energy. The lines are polynomial fits to
 data.  
}
\end{figure}

The Fermi statistics of  $^3$He atoms is the key point underlying the
non-self-bound character of  2D $^3$He. This conclusion is drawn
from the comparison between the real system and a fictitious 2D $^3$He boson
system. In Fig. 3, the DMC equation of state of both systems is compared.
Boson $^3$He would show a liquid phase with a binding energy $(E/N)_0=-0.1189(23)$ K
at an equilibrium density $\rho_0=0.1311(17)\sigma^{-2}$. 
The lighter mass
of the $^3$He atom is responsible for the large reduction of binding
energy of boson $^3$He
with respect to liquid $^4$He: using the same potential, the $^4$He
equilibrium point is ($0.284\sigma^{-2},-0.897$K)~\cite{boro2d}. Some previous
theoretical calculations on 2D $^3$He used the boson model as a reference
system~\cite{novaco,nosanow}. On top of this, dominant Fermi corrections 
were added to introduce
the correct statistics. This perturbative approach, known as Wu-Feenberg
expansion~\cite{wu}, 
starts at zeroth order by simply adding $E_{\text
F}/N$ to the boson energy. This crude estimation is plotted in Fig. 3. As
one can see, the kinetic term $E_{\text F}/N$  
leads to a large overestimation of $E/N$, and
only for $\rho < 0.03\sigma^{-2}$ can it be considered a reasonable
approximation. The successive terms in the Wu-Feenberg series show a non monotonic
behavior and in general a very slow convergence: at medium densities ($\rho
\simeq 0.1\sigma^{-2}$) the next-order term is negative and approximately 
100 times
smaller than the zeroth order~\cite{chester2d}. Therefore, although
the present results modify quantitatively those approximate calculations,
previous conclusions about the gas-like character of $^3$He are not
altered.

Relevant quantities from the experimental standpoint are the density
dependence of the pressure $P(\rho)$ and the speed of sound $c(\rho)$.
Both functions are shown in Fig. 4. They have been obtained from the  
polynomial fit to the DMC energies (\ref{eqestatpol}) through the
thermodynamic relations $P(\rho)=\rho^2 ( d(E/N)/d \rho)$ and $c(\rho)=[
m^{-1} (d P/ d \rho) ]^{1/2}$.  The pressure remains very low up to
$\rho \simeq 0.20\sigma^{-2}$ and from then on increases much faster
due to the small $^3$He mass and the rapid decrease with density of the 
mean distance
between particles,  due to the reduced dimensionality. 
An approximate estimate for the
latter comes from $2 (4 \pi \rho)^{-1/2}$; at $\rho=0.25\sigma^{-2}$ it
amounts only  2.88\AA, a smaller value than 
in bulk $^3$He at freezing, 4.32\AA. The speed of sound presents
three different regimes as a function of $\rho$. At very small densities,
$c(\rho)$ increases approximately like $\rho^{1/2}$ as it would correspond
to a free 2D Fermi  gas. Then, $c(\rho)$ reaches a plateau up to $\rho
\simeq 0.08\sigma^{-2}$. In this region, the speed of sound increases 
although the slope is very small;
this behavior is a direct consequence of the flattening exhibited by the 
energy for the same range of densities (see Fig. 2).
Finally, in the third regime $c(\rho)$ again increases with $\rho$ in a
more common way.

\begin{figure}
\centering
\includegraphics[width=7cm]{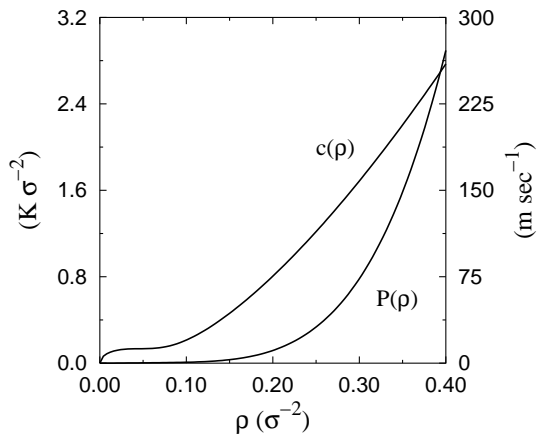}
\caption{\label{fig:fig4} Density dependence of the pressure and  speed
of sound of 2D $^3$He. 
}
\end{figure}

The 2D system constitutes a model for a film adsorbed on a
strong substrate like graphite. A question that naturally emerges is to
what extent the presence of a transverse degree of freedom would modify 
the $^3$He properties. First, the nodal surface could be
different from the one of the 2D system. However, a key result of this work 
in 2D,
and the previous one in 3D~\cite{bulk3}, is that the real  nodal surface 
in both cases is essentially given by backflow  effects. 
In a film, the  particle-particle backflow correlations would be mainly
contained in the surface plane. Furthermore, both the not-in-plane 
contributions and the particle-substrate correlations, being perpendicular
to the surface plane, should have a small effect on the backflow wave function.
%In a film, these dynamical correlations
%would largely be contained in the surface plane, thus originating a final
%nodal surface almost identical to that of a purely 2D system. 
Therefore,
the Fermi statistics of a thin $^3$He film  can be safely considered within
the idealized 2D geometry.  
Second, the additional
degree of freedom perpendicular to the substrate could by itself lower
the energy in an amount large enough to allow for the existence of a
liquid phase.
In fact, a VMC calculation of $^3$He and $^4$He films adsorbed on graphite 
by Brami \textit{et
al.}~\cite{brami} concludes for the former the existence of a self-bound 
system with a binding energy
of $\sim 200$mK at an equilibrium density $\sim 0.131\sigma^{-2}$. To our
knowledge, there are not DMC or GFMC calculations of $^3$He films on
graphite that can confirm that variational prediction. However, Whitlock
\textit{et al.}~\cite{whitlock2} performed GFMC calculations for $^4$He films 
adsorbed on the same
substrate, and found a decrease in energy with respect to 2D that is much
smaller than the results from Ref.~\cite{brami}. 
Even at the variational level,
and using the same wave function and graphite-helium potential, Whitlock 
\textit{et al.}~\cite{whitlock2}
were not able to reproduce the $^4$He  energy reported in Ref.
\cite{brami} (-0.7 vs. -1.9K ). 

The order of magnitude of the shift in energy that appears in a $^3$He 
film with respect to the 2D system can be estimated from the GFMC results 
for the $^4$He liquid~\cite{whitlock2}, taking into account 
the different mass of the two isotopes in the approximate kinetic-energy correction. 
For example, at densities
$\rho=0.065\sigma^{-2}$, $0.131\sigma^{-2}$, and $0.170\sigma^{-2}$ the
energy shifts are  $\Delta E/N= -0.005$K, $-0.022$K, and $-0.037$K, to be
compared with the 2D energies $E/N=0.100$K, $0.159$K, and $0.220$K,
respectively. Therefore, the energy shift is by far too small to change the 
conclusion that, like the strictly 2D fluid, a thin $^3$He film is not 
self-bound.

We acknowledge financial support from DGES
(Spain) Grant No. PB98-0922 and Generalitat de Catalunya Grant No.
2001SGR-00222. Supercomputer facilities provided by CEPBA are also
acknowledged.

%\begin{figure}
%\centering
%\includegraphics[width=6cm]{grdens.eps}
%\caption{\label{fig:fig1}See what beautiful is}
%\end{figure}

\end{document}